\documentstyle[prl,aps,twocolumn]{revtex}

\begin{document}

\draft
\title{Comment on ``Jellium model of metallic nanocohesion''}
\author{C. H\"oppler and W. Zwerger}
\address{Sektion Physik, Ludwig-Maximilians-Universit\"at,
Theresienstra{\ss}e 37, D-80333 M\"unchen, Germany}
\date{\today}
\maketitle

In a recent letter Stafford, Baeriswyl and B\"urki \cite{stafford} have shown
that the elastic force in metallic constrictions of nanometer
size can be understood within a simple model, in which the
discrete transverse eigenstates of the conduction electrons act
as delocalized bonds providing cohesion. This model qualitatively
accounts both for the size of the average force as well as for the
oscillations of order $\varepsilon_F/\lambda_F$ due to 
switching off successive transverse modes, which is also
responsible for the simultaneously measured steps in the 
conductance \cite{rubio}. Here we would like to point out that the two
simple models for the cross section used in Ref. \cite{stafford} actually
give different results for the average force. This explains the
apparent discrepancy in the offsets of their Fig. 3 showing the
force oscillations. Quite generally, for any smooth cross
section, there is a universal contribution $4\varepsilon_F/
9\lambda_F$ to the average force which is absent in the presence
of a hole. 

In the theory of Ref.\cite{stafford} the force $F=-\partial\Omega/\partial L$
is obtained from the grand canonical potential $\Omega$ of
noninteracting electrons at zero temperature. The restriction to the kinetic
energy of the electrons requires that exchange and correlation contributions as
well as the attraction to the ions remain unchanged under a deformation at 
constant volume. In this case $\Omega$ can be written as
\begin{equation}
\Omega(T=0,L)=-\frac{4}{\lambda_F}\int_0^L\!\!\!dz\int_0^{\varepsilon_F}
\!\!\!\!\!\!dE\, N_{\perp}(E,z)\left(1-\frac{E}{\varepsilon_F}\right)^{1/2},
\end{equation}
with $N_{\perp}(E,z)$ the integrated density of states of the
discrete transverse modes at a given cross section $z=$const. 
This function may be split into a smooth average $\bar N_{\perp}
(E,z)$ and an oscillating piece $\delta N_{\perp}$ which is 
related to the classical periodic orbits. In the semiclassical
limit, the average number of transverse modes at any given $z$
which have energies smaller than $E$ is obtained from an
extended Weyl expansion \cite{kac}, which reads
\begin{equation}
\bar N_{\perp}(E,z)=\frac{A(z)}{4\pi}k_E^2-\frac{\partial A(z)}
{4\pi}k_E+\frac{1}{6}(1-p)+\ldots\, .
\end{equation}
in the case of Dirichlet boundary conditions.
Here $k_E=\sqrt{2ME/\hbar^2}$ is the wave number
associated with energy $E$ while $A(z)$ and $\partial A(z)$
denote the area and circumference of the cross section. The 
third contribution is a universal term of purely topological
origin which only depends on the number $p=0,1,\ldots$ of holes
in the cross section. The result (2) applies to {\it arbitrary}
smooth boundaries but not in the case of corners. Taking a
square cross section for instance as in Ref. \cite{stafford}, the topological
contribution $1/6$ is replaced by $1/4$. Inserting the latter 
value into (1), one immediately obtains Eq. (11) in Ref.\cite{stafford}.
For the more physical case of a smooth boundary however - e.g. 
a circle which was also considered in Ref. \cite{stafford} - 
the coefficient
of the contribution to $\bar\Omega$ which is linear in $L$ is
different by a factor $\frac{2}{3}$.

Considering deformations at constant total volume $V=\int_0^Ldz\, A(z)$,
the semiclassical expansion 
(2) implies that the average cohesive force of an arbitrary
smooth constriction
\begin{equation}
\bar F=-\frac{\partial\bar\Omega}{\partial L}=
-\frac{\varepsilon_F}{\lambda_F}\left[\frac{k_F}{8}
\left.\frac{\partial S}{\partial L}\right|_V\, -\frac{4}{9}(1-p)
+\ldots\right]
\end{equation}
has a universal contribution which is sensitive to the topology
of the cross section. In particular this term, which weakens the
cohesive force compared to the macroscopic surface tension 
contribution, is absent in the presence of a hole, i.e. for a
constriction with the topology of a hollow cylinder. Since the
change in surface area $S=\int_0^Ldz\,\partial A(z)$ with 
length $L$ at constant volume is of the order of the constriction
radius $R$ (one has $\frac{\partial S}{\partial L}\vert_V=\pi R$
for an ideal cylinder), the terms in the square bracket in (3)
are the two leading contributions in a semiclassical expansion
with $k_FR\gg 1$. In this limit the corrections indicated by the
dots are negligible. As a result, the contributions from the
fluctuating term $\delta N_{\perp}$ to the density of states
give rise to force fluctuations $\delta F$ which vanish if
averaged over sufficiently many modes. This is verified
explicitely in Figs. 3b and c of Ref. \cite{stafford}, where $\delta F$ is
calculated for the square geometry in which Eq. (12) in Ref. \cite{stafford}
is correct. By contrast, for the circular cross section shown
in Fig. 3a, it is straightforward to check that a vanishing 
average only results if the curve is shifted upwards by
$2\varepsilon_F/9\lambda_F$, thus accounting for the proper
topological term as given in Eq. (2) above ($p=0$). It would
obviously be of considerable interest, to verify experimentally
the increase in the cohesive force at given $V$ and 
$\partial S/\partial L$ which follows from Eq. (3)
for constrictions with a hole in the cross section.

\end{document}